\documentstyle[12pt]{article}
\begin{document}
\noindent{\Large \bf Evidence of potential sputtering of ultra-thin Pt films
due to impact of multi-charged Ar ions} \vspace{.7in}\\
\begin{center}
{\bf D. Ghose\footnote{Corresponding author. e-mail: ghose@cmp.saha.ernet.in} and P. Karmakar} \vspace{.1in} \\
{\it Saha Institute of Nuclear
Physics,\\ Sector - I, Block - AF, Bidhan Nagar, Kolkata 700064, India} \vspace{.4in}\\
\today \vspace{.2in}\\
{\bf Abstract}\\
\end{center}

The phenomenon of potential sputtering of  ultra-thin Pt films due to impact of 20 keV highly charged  $Ar^{q+}$ (q = 3 - 8) ions has been investigated. The results show enhanced erosion of the Pt film  with the increase of  potential energy of the Ar projectile.\\

\pagebreak

In recent years the interactions of multiply charged ions with surfaces have been the subject of active research both from intrinsic interest to understand various emission processes as well as future technological applications [1 - 3]. For highly charged ions, the potential energy stored in the projectile can exceed far greater than the projectile's kinetic energy. In contrast to the well-known kinetically induced ion beam sputtering process, a new phenomenon called potential sputtering, i.e. the erosion of surfaces exploiting the potential energy of the projectile has recently been observed [4, 5]. In this communication, we report on first time the observation of enhanced sputtering of ultra-thin Pt films when bombarded by highly charged Ar ions.\\
 
Pt thin films of $\simeq$ 3 nm thickness were deposited by DC-magnetron sputtering onto commercially available polished Si (100) wafers, previously degreased and cleaned. The base pressure in the deposition chamber was $2\times10^{-6}$ mbar. The sputtering experiments were performed using the PANTECHNIK s.a. 14.5 GHz Electron Cyclotron Rsonance (ECR) ion source recently installed at the Variable Energy Cyclotron Center , Kolkata . Ar ions of different charge states, q (= 3 - 8), were incident on the samples at an angle of $45^0$ with the surface normal. The residual gas pressure in the target chamber was $5\times10^{-8}$ mbar. The current density was varied in the range 1.1 - 5.6 $\mu A/cm^2$ depending on the charge state of the projectile species. The ion dose was measured by a Danfysik model 554 current integrator after suppression of the secondary electrons. The irradiated samples were analyzed by Rutherford backscattering spectroscopy (RBS) using a 2 MeV $He^{2+}$ beam at a scattering angle of $110^0$. Each spectrum was normalized with the same ion fluence and was evaluated using PC-based GISA-3.991 program [6].\\

Figure 1 shows the superposition of five backscattering spectra of Pt from an unbombarded sample and the samples bombarded with the projectiles of charge states, q = 3, 4, 5 and 8, respectively. The kinetic energy of the projectile species was kept constant at 20 keV and each sample was bombarded with a fixed dose of $1\times10^{16}$ particles/$cm^2$. The most interesting observation is that the area under the Pt peak decreases steadily with increasing charge state, which is a clear signature of potential sputtering of the Pt film. Previously, such a phenomenon was observed for insulator targets, e.g., LiF [7], $SiO_2$ [8] and more recently in $MgO_x$ [9].\\
 
There exist mainly two competing models of potential or electronic sputtering, namely, "Coulomb explosion (CE)" [10] and "defect-mediated sputtering (DS)" [7]. In the CE model, the positive target ion cores in the upper most layers are exposed due to the rapid electron capture into the incoming highly charged ions. In insulators and poor conductors, the charge neutrality can not be reestablished on the time scale of lattice vibrations. As a consequence, the target particles are exploded out due to the mutual electrostatic Coulomb repulsion to an amount approximately proportional to the potential energy carried by the projectile ion. On the other hand, the DS model lends itself the formalisms developed for electron- and photon-stimulated desorption phenomena in alkali halide crystals. It considers the formation of localized defects such as self-trapped excitons in response to valence band excitations. These defects subsequently decay into colour centers which may diffuse to the surface and lead to the desorption of target atoms. For heavy and very highly charged ions which result high-density electronic excitations at the target surface, a third mechanism of potential sputtering was invoked [11], which involves structural instabilities arising from destabilization of atomic bonds.\\

In the literature, e.g. [5], there is a continuous debate about different erosion mechanisms and no general consensus has yet been achieved. The defect-mediated desorption model is shown to be operative for some selected ionic materials where strong electron-phonon coupling exists. It is argued that the trapping of electronic excitations by forming defects is a prerequsite for the conversion of potential energy of the projectile into kinetic energy of the target particles. Recently, Hayderer et al. [9] proposed a new mechanism of potential sputtering, namely, "kinetically assisted potential sputtering" for the materials where electron-lattice coupling may not have to be strong enough. In this case trapping of excitons is possible at Frenkel defects or higher order defects caused due to momentum transfer by the projectile ion. The possibility of localization of electronic excitations in the collision cascade zone is also evident from some typical experimental results of ionization and excitations in sputtering process as discussed by \v{S}roubek and L\"orin\v{c}\'ik [12]. In order to explain those results, these authors assumed that the excitations formed within the collision cascade volume do not dissipate immediately after the ion impact but can survive for a sufficiently long time, even in metals, because of the destruction of the local lattice structure by the bombarding ion. For example, a rough calculation shows that the d-band holes in Ag trapped in a lattice disorder after violent collisions by 20 keV $Ar^+$ ions can survive as long as 100 fs [13]. Following the works of Knotek [14], Mochiji et al. [15] have pointed out that the holes can also be localized at the creation site with extended lifetimes ($ > 10^{- 14}$ sec) when many holes are generated at close vicinity to each other.\\ 

It is well known that the resistivity of metallic thin films increases over the bulk resistivity as the film thickness decreases and finally, approaches the resistivity of the underlying substrate [16]. For example, it has been found [17] that for a 3 nm Pt film the resistivity is a factor of $\sim$16 higher than that for bulk material. Thus in the present experiment, the ultra-thin Pt film is thought to be a bad conductor for which the surface Coulomb explosion model may be applied. In this context it may be mentioned that Stiegler and Noggle [18] noticed a similar type of ion explosion process responsible for the development of ion tracks in very thin Pt films following impact of fast nitrogen ions.\\

In order to calculate the sputtering yield, one has to determine the thickness of the eroded film. Since the nominal thickness of the Pt film is much smaller than the depth resolution (15 - 20 nm) of the RBS, a direct analysis of the Pt areal density is not possible. However, an estimate of the erosion yield can be made from the integrated area of the observed RB spectra by normalizing the area corresponding to zero neutralization energy with the "kinetic" sputtering yield, $Y_{KS}$, which can be calculated by the SRIM-2000 code [19]. In Fig. 3 the total sputtering yield, $Y$, as obtained from RBS data are shown as a function of the potential energy of the projectile, $W$. $Y$ consists of two components, one due to potential sputtering ($Y_{PS}$) and another due to kinetic sputtering ($Y_{KS}$). The CE-model [10] predicts that the $Y_{PS}$ is proportional to the volume of the charged domain, i.e. $(R - a)^3$, where $R$ is the radius of the multiple hole hemisphere and is dependent on the material property and the potential energy of the projectile. $a$ is the thickness of the domain border shell quenched during the neutralization time and varies inversely with the the lifetime of the holes, $\tau$. A best fit to the experimental data was obtained when $a \to 0$. That means that the lifetime of the holes in the charged domain is considerably high. For proton sputtering from a hydrocarbon surface induced by highly charged ions, Burgd\"orfer and Yamazaki [20] estimated $\tau$ lying in the range $10^{-14} - 10^{-15}$ sec. Following the analysis of Bitenskii et al. [10], the present data can, thus,  be described by the relation:
\begin{equation} 
Y = k.W^{3/5} + Y_{KS},
\end{equation}
where the material constant $k$ is found to be about 0.26. This value is quite reasonable since a small fraction ($\sim$10\%) of the initial potential energy is available for the emission of the secondary particles [2].\\

In summary, the present experiment shows that the erosion of ultra-thin metal film is greatly enhanced by the potential energy of the projectile ion. The results exhibit a $W^{3/5}$ dependence on the potential energy corresponding to the Coulomb explosion model of sputtering.\\

The authors thank Dr. S. Kundu for the preparation of the Pt/Si samples, Mrs. P. Agarwal, Mr. P. Y. Nabhiraj and Dr. D. K. Bose for their involvement in the irradiation with the ECR ion source. The RBS work was carried out with the 3 MV tandem Pelletron accelerator at IOP, Bhubaneswar.

\pagebreak

\section*{References}
\begin{itemize}

\item[[1]] A. Arnau, F. Aumayr, P. M. Echenique, M. Grether, W. Heiland, J. Limburg, R. Morgenstern, P. Roncin, S. Schippers, R. Schuch, N. Stolterfoht, P. Verga, T. J. M. Zouros and HP. Winter, Surf. Sci. Rep. {\bf27}, 113 (1997).
\item[[2]] T. Schenkel, A. V. Hamza, A. V. Barnes and D. H. Schneider, Prog. Surf. Sci. {\bf61}, 23 (1999).
\item[[3]] J. D. Gillaspy, J. Phys. B {\bf34}, R93 (2001).
\item[[4]] F. Aumayr, J. Burgd\"orfer, P. Varga and HP. Winter, Comments At. Mol. Phys. {\bf34}, 201  (1999).
\item[[5]] T. Schenkel, M. W. Newman, T. R. Niedermayr, G. A. Machicoane, J. W. McDonald, A. V. Barnes, A. V. Hamza, J. C. Banks, B. L. Doyle and K. J. Wu, Nucl. Instr. and Meth. B {\bf{161 - 163}}, 65 (2000).
\item[[6]] J. Saarilahti and E. Rauhala, Nucl. Instr. and Meth. B {\bf64}, 734 (1992).
\item[[7]] T. Neidhart, F. Pichler, F. Aumayr, HP. Winter, M. Schmid and P. Varga, Phys. Rev. Lett. {\bf74}, 5280 (1995).
\item[[8]] M. Sporn, G. Libiseller. T. Neidhart, M. Schmid, F. Aumayr, HP. Winter, P. Varga, M. Grether, D. Niemann and N. Stolterfoht, Phys. Rev. Lett. {\bf79}, 945 (1997). 
\item[[9]] G. Hayderer, S. Cernusca, M. Schmid, P. Varga, HP. Winter, F. Aumayr, D. Niemann, V. Hoffmann, N. Stolterfoht, C. Lemell, L. Wirtz and J. Burgd\"orfer, Phys. Rev. Lett. {\bf86}, 3530 (2001). 
\item[[10]] I. S. Bitenskii, M. N. Murakhmetov and E. S. Parilis, Sov. Phys. Tech. Phys. {\bf24}, 618 (1979).
\item[[11]] T. Schenkel, A. V. Hamza, A. V. Barnes, D. H. Schneider, J. C. Banks and B. L. Doyle, Phys. Rev. Lett. {\bf81}, 2590 (1998).
\item[[12]] Z. \v{S}roubek and J. L\"orin\v{c}\'ik, Surface Review and Letters {\bf6}, 257 (1999).
\item[[13]] A. Wucher and Z. \v{S}roubek, Phys. Rev. B {\bf55}, 780 (1997).
\item[[14]] M. L. Knotek, Phys. Today (Sept. 1984) 2.
\item[[15]] K. Mochiji, S. Yamamoto, H. Shimizu, S. Ohtani, T. Seguchi and N. Kobayashi, J. Appl. Phys. {\bf82}, 6037 (1997).
\item[[16]] S. K. Song, H. J. Jung, S. K. Koh and H. K. Baik, Appl. Phys. Lett. {\bf71}, 850 (1997).
\item[[17]] M. Avrekh, O. R. Monteiro and I. G. Brown, Appl. Surf. Sci. {\bf158}, 217 (2000).
\item[[18]] J. O. Stiegler and T. S. Noggle, J. Appl. Phys. {\bf33}, 1894 (1962).
\item[[19]] J. F. Ziegler, TRIM code, version SRIM-2000; described at http://www.research.ibm.com
\item[[20]] J. Burgdo\"rfer and Y. Yamazaki, Phys. Rev. A {\bf54}, 4140 (1996).

\end{itemize}  

\pagebreak

\section*{Figure captions}

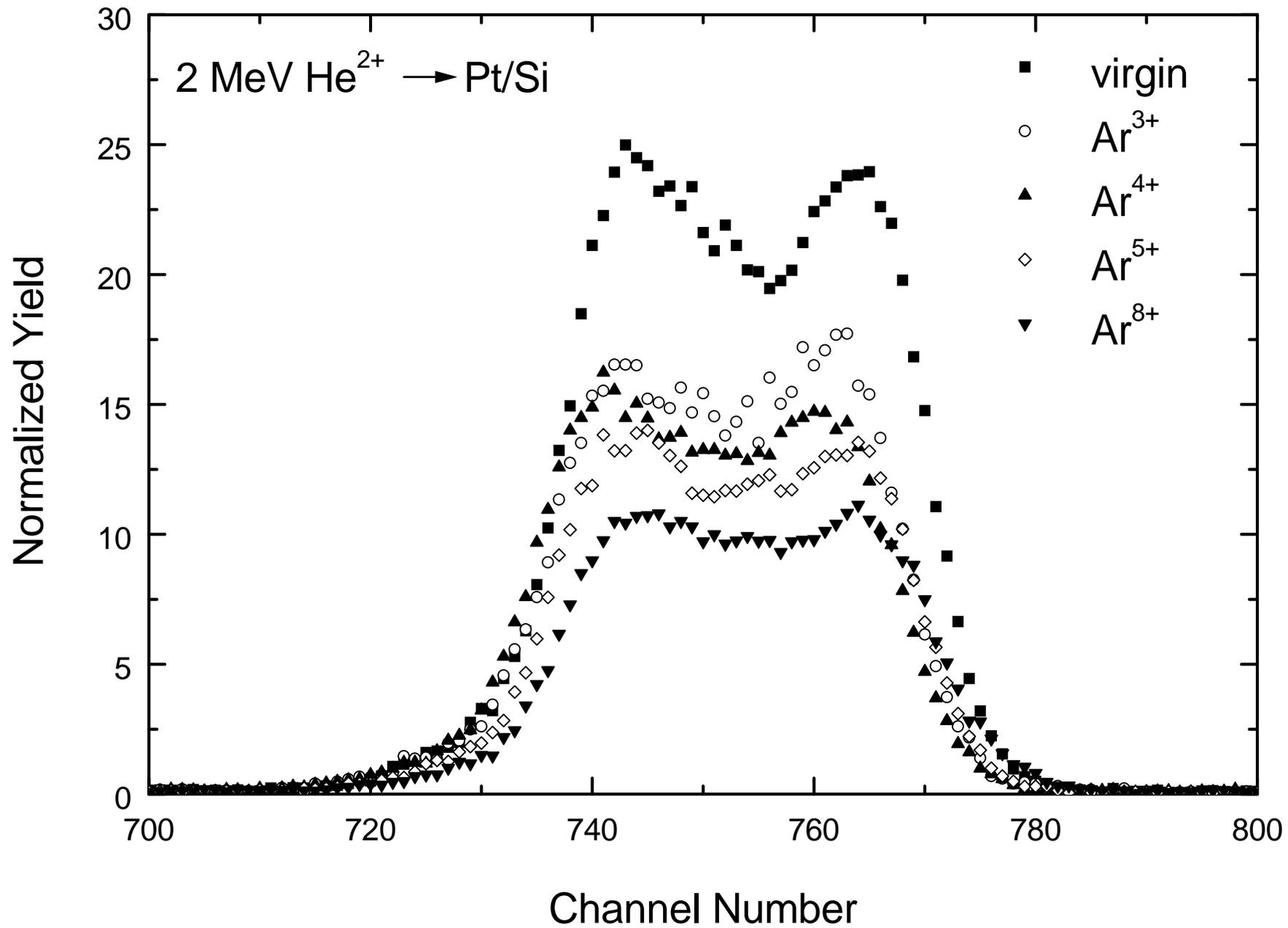
Fig. 1. Superimposed backscattering spectra of Pt from irradiated Pt/Si samples at various charge states of the incident Ar ion.\\

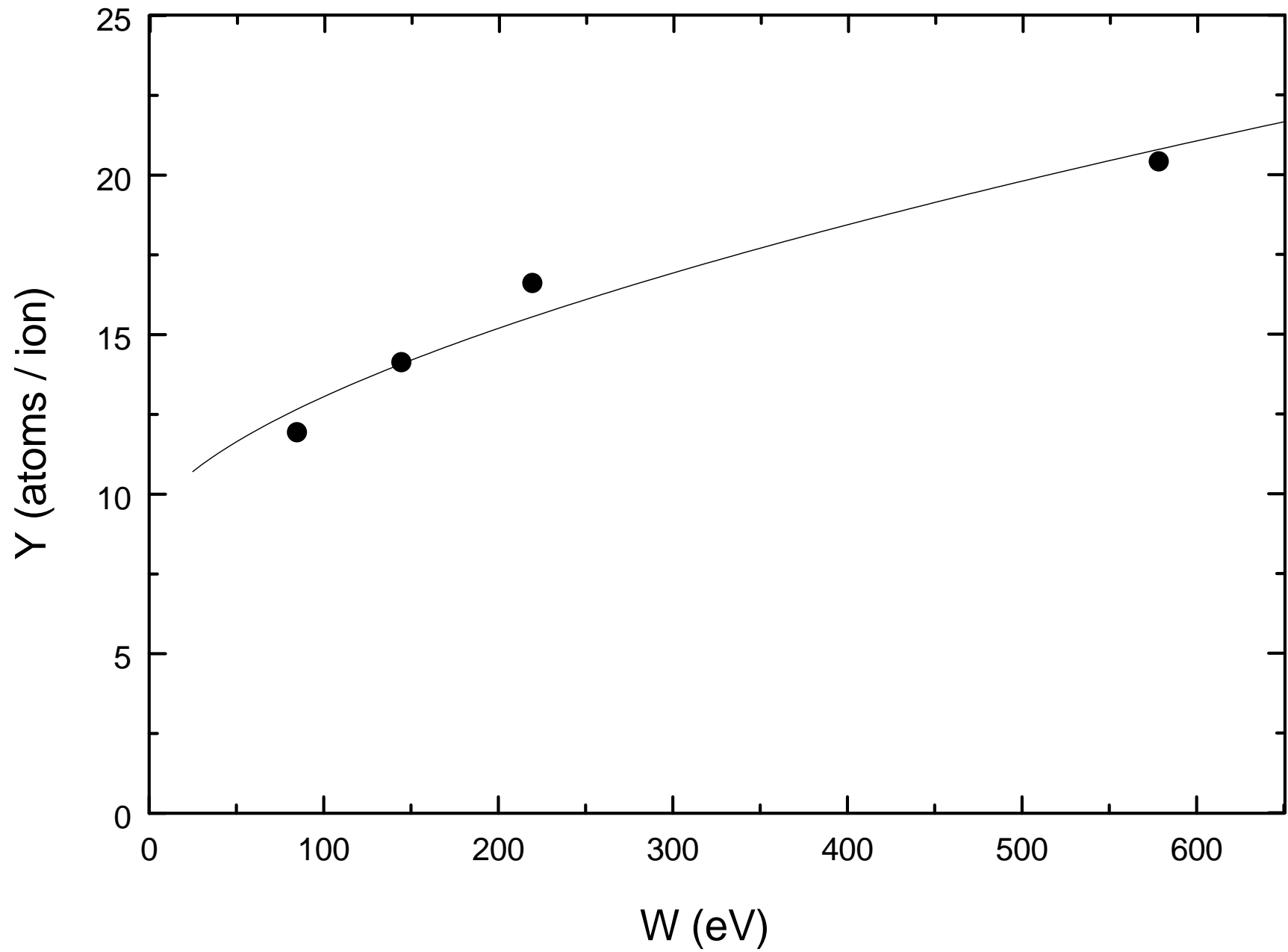
Fig. 2. Potential energy dependence of the erosion yield of Pt. $\bullet$ - experimental data points; solid line - CE model.\\

\end {document}